\documentstyle[12pt,aasms4]{article}

\received{}
\accepted{}
\journalid{}{}
\articleid{}{}
\lefthead{Hatano et al.}
\righthead{Spatial Distribution of M31 Novae}

\begin{document}
\title{New Insight into the Spatial Distribution of Novae in M31}

\author{KAZUHITO HATANO, DAVID BRANCH, ADAM FISHER\altaffilmark{1}}

\affil{Department of Physics and Astronomy, 
University of Oklahoma, Norman, OK 73019}

\altaffiltext{1}{Department of Physics, Southwestern Oklahoma State
University, Weatherford, OK 73096}

\author{SUMNER STARRFIELD}
\affil{Department of Physics and Astronomy, Arizona State
University, Tempe, AZ 85287} 

\begin{abstract}

We use a Monte Carlo technique together with a simple model for the
distribution of dust in M31 to investigate the observability and
spatial distribution of classical novae in M31.  By comparing our
model positions of novae to the observed positions, we conclude that
most M31 novae come from the disk population, rather than from the
bulge population as has been thought.  Our results indicate that the
M31 bulge--to--disk nova ratio is as low as, or lower than, the M31
bulge--to--disk mass ratio.

\end{abstract}

 
\section{Introduction}

Opinions about the spatial distribution of classical novae in M31 and
in our Galaxy have been undergoing an interesting evolution. The major
searches for novae in M31 (Hubble 1929; Arp 1956; Rosino 1964, 1973;
Rosino et al. 1989) showed that in a general sense novae are
distributed like the light of the galaxy, apart from a possible
deficit of novae (a ``nova hole") within the central few minutes of
arc.  By means of a CCD H$\alpha$ survey, Ciardullo et al. (1987) found
novae in the innermost regions and concluded that the nova hole was
just due to incompleteness caused by saturation on the photographic
plates that had been used in the earlier nova surveys.  Ciardullo et
al. also concluded that novae in M31 belong overwhelmingly to the
bulge population.  Capaccioli et al. (1989) reached the same
conclusion.  At least partly because of the belief that M31 novae are
overwhelmingly from its bulge population, it is often assumed that
Galactic novae also come mainly from the bulge; for example, Della
Valle \& Duerbeck (1993) and Della Valle \& Livio (1994) assumed that
3/4 of the Galactic novae are from the bulge.

Doubts about the bulge dominance of the M31 nova population arose when
Ciardullo et al. (1990) combined the results of their search for novae
in NGC 5128 with data on novae in the LMC, SMC, M33, M31, and a few
elliptical galaxies in the Virgo cluster.  They found the nova rates
per unit K--band luminosity to be remarkably similar --- apart from a
strikingly low rate for the M31 disk.  They suggested that since the
M31 bulge has been more thoroughly searched for novae than its disk,
and since disk novae may be preferentially obscured by dust, it is
possible that the nova rate in the M31 disk had been underestimated,
and that the nova rate per unit mass of old stellar population may be
approximately a constant.  Recently, Shafter, Ciardullo, \& Pritchet
(1996) report that their search for novae in three more galaxies, M51,
M101, and M87, has produced preliminary results that are consistent
with this proposition.

Another point of view that has developed recently is that young
populations are {\sl better} than old populations at producing
novae. On the basis of observation, Della Valle et al. (1994)
concluded that bulge--dominated galaxies (NGC 5128, M31, M81, and
Virgo ellipticals) have a nova rate per unit H--band luminosity that
is more than a factor of three lower than that of nearly bulgeless
galaxies (LMC and M33).  And, on the basis of a binary--star
population--synthesis study, Yungelson, Livio, \& Tutukov (1997)
predict that the nova rate per unit mass of a young population should
be much higher than that of an old population.  Yungelson et al. find
support for that prediction in the nova rates per unit K--band
luminosity in the galaxies mentioned above, and they suggest that the
apparent dominance of bulge novae in our Galaxy may be due to
observational selection effects that favor the discovery of bulge
novae over disk novae.

Recently we (Hatano et al. 1997) have used a Monte Carlo technique
together with a simple model for the distribution of dust in the
Galaxy to investigate the observability and spatial distribution of
Galactic classical novae.  We concluded that most Galactic novae are
indeed produced by the disk, rather than by the bulge.  More
specifically, we found the distribution of nova apparent magnitudes
and positions on the sky to be consistent with the proposition that
the Galactic bulge--to--disk nova ratio is equal to that of the
overall Galactic bulge--to--disk mass ratio, which is only about 1/7
(van den Kruit 1990).  In this Letter we report results of a study
which set out to address the question of whether, similarly, the M31
bulge--to--disk nova ratio is consistent with the M31 bulge--to--disk
mass ratio, which is about 1/2 (Kent 1989; Hodge 1992).

\section{Observations}
 
Since about 1917, more than 300 novae have been discovered in M31.  We
concentrate on 191 novae that were discovered (or reported) in major
surveys carried out at Mount Wilson (Hubble 1929; Arp 1956) and at
Asiago (Rosino 1964, 1973; Rosino et al. 1989), and for which
estimates of the peak apparent visual magnitude, $V$, are available.

Fig. 1 shows the positions on the sky of these novae, on the
coordinate system of Capaccioli et al. (1989).  At the adopted
distance to M31 of 725 kpc ($\mu=24.3$), six arcminutes corresponds to
about 1 kpc.  We take the bulge to be spherical, with a radius of 18
arcminutes, or three kpc.  Obviously, most of these observed novae are
projected within the bulge.  The major--to-minor axis ratio of M31 is
4.3, for an inclination of 77 degrees.  The ellipse in Fig. 1
corresponds to a circle in the disk, of radius 8.8 kpc (where, as
described below, the density of the dust peaks in our model).  Note
that a great deal of disk is projected within the adopted perimeter of
the bulge.

We define ``apparent bulge" novae to be those whose sky positions are
within the 18--arcminute radius of the bulge, and ``apparent disk
novae" to be those whose positions are not.  As discussed below, some
of the apparent bulge novae actually are disk novae.  The top panel of
Fig. 2 shows the $V$--distributions for the 176 apparent bulge novae,
the 15 apparent disk novae, and the sum of the two.  (For comparison,
the shape of our model $V$--distribution, to be discussed below, also
is shown in the top panel.)

\section{The Model}

The Monte Carlo technique that we have developed was inspired by one
that was used by Dawson \& Johnson (1994) in an interesting study of
the observability of historical supernovae in our Galaxy.  We (Fisher
et al. 1997) constructed an independent Monte Carlo code and used it
to extend the work of Dawson and Johnson by considering the
observability of hypothetical ``ultra--dim" supernovae in the Galaxy,
and to consider the observability of supernovae, in the model, from an
external point of view.  Then we (Hatano et al. 1997) extended the
technique to consider the observability of Galactic classical novae.
Here we give a brief description of the model as it is used for this
study of novae in M31.

In our previous papers, the Galactic dust was assumed to be
distributed according to a simple double exponential law, with a
radial scale length of 5 kpc and a vertical scale height of 0.1 kpc.
In such a model, the density of the dust peaks right at the center of
the galaxy.  In M31, however, the density of the dust is known to peak
well out in the disk, not far from where most of the current star
formation rate is taking place (Hodge 1992).  Following Fig. 3 of Xu
\& Helou (1996), we adopt a simple distribution for the radial
dependence of the extinction in M31:

$$ A_V = 2.0 - 0.182 (8.8 - r),\ \ for\ \  r < 8.8 kpc, \eqno (1) $$
 
$$ A_V = 2.0 - 0.194 (r - 8.8),\ \ for\ \  r > 8.8 kpc, \eqno (2) $$

\noindent where $A_V$ is the total line--of--sight extinction through
the inclined disk of M31.  This distribution is generally consistent
with the various evidence for the radial dependence of extinction
discussed by Hodge (1992).  The vertical scale height of the dust is
taken to be 0.1 kpc, as we used for our Galaxy.  In this model, the
extinction at $r=8, z=0$ kpc is 1.85 mag kpc$^{-1}$, similar to its
value at $r=8, z=0$ kpc in our Galactic model, 1.9 mag kpc$^{-1}$.
The major difference between our adopted distributions of dust in the
Galaxy and in M31 is the low dust content in the central regions of
M31.

Disk and bulge novae in M31 are assigned the same spatial
distributions as we used for our Galaxy.  Disk novae obey a double
exponential distribution, with radial and vertical scale lengths of 5
and 0.35 kpc, and the disk is truncated at $r=20$ kpc. Bulge novae
are taken to be distributed as $(R^3 + a^3)^{-1}$, where $R$ is a
radial coordinate, $R^2 = r^2 + z^2$ , and $a=0.7$ kpc.  The bulge is
truncated at $R=3$ kpc, and the disk and bulge components
interpenetrate.  For Galactic novae, we used a bulge--to--disk nova
ratio of 1/7, based on the estimated bulge--to--disk mass ratio of the
Galaxy (van der Kruit 1990).  For M31, a more reasonable estimate of
the bulge--to--disk mass ratio would be 1/2 (Hodge 1992; Kent
1989), so we adopt this as our default value of the M31
bulge--to--disk nova ratio.

The M31 nova luminosity functions also are taken to be the same as we
used for novae in our Galaxy.  They are gaussian, with dispersions
$\sigma(M_V)=1$, and the mean absolute magnitudes of disk and bulge
novae are $-8$ and $-7$, respectively.

\section {Comparison with Observations}

The middle panel of Fig. 2 shows our model $V$--distributions for true
bulge novae, true disk novae (we do know which {\sl model} novae are
from the disk and which are from the bulge), and the sum of the two.
As can be seen in the top panel of Fig. 2, the total model
$V$--distribution agrees well with the observed $V$--distribution on
its bright side, to $V \simeq 16.5$.  However, the model distribution
contains a larger proportion of faint novae than the observed
distribution.  This is due at least in part to observational selection
against faint novae (many faint observed novae had to be excluded from
our sample because no estimate of peak apparent magnitude was
available), but it may also be that our adopted luminosity functions
contain too many intrinsically dim novae; in any case this will not
affect our main conclusion because it will be based only on the
brighter novae. Note that in the mean, the true disk novae are
brighter than the true bulge novae.  The bottom panel of Fig. 2 is
like the middle one, except that now the model novae are divided into
{\sl apparent} disk and {\sl apparent} bulge novae, on the basis of
whether or not their projected positions are within 18 arminutes of
the center of M31. Because of the presence of true disk novae
masquerading as apparent bulge novae, the difference between the
$V$--distributions of the apparent disk and apparent bulge novae is
smaller than the difference between the $V$--distributions for the true
disk and true bulge novae.  In the middle panel of Fig. 2, almost all
of the bright novae are true disk novae, but in the lower panel, many
of those true disk novae become apparent bulge novae.  And, even
though our input model bulge--to--disk nova ratio is only 1/2, the
apparent bulge novae outnumber the apparent disk novae.  Therefore,
according to our model, {\sl a large fraction of the apparent bulge
novae in M31 actually are disk novae}.

Some insight into what is going on (at least in the model) can be
gained from Fig. 3, which shows a side view of the spatial
distribution of model novae having $V < 20$; for clarity, the vertical
scale is expanded by a factor of five.  First, many true disk novae
having $r < 13.9$ kpc are seen as apparent bulge novae.  Second, while
true bulge novae on the top side of the bulge are practically
unextinguished, from our vantage point, true bulge novae on the bottom
are significantly extinguished by dust that is well out in the disk,
where the extinction is largest.  This means that {\sl true bulge
novae projected onto the top of the bulge are, in the mean, brighter
than those projected onto the bottom.}  As can be inferred from
looking at Fig. 3, the $V$--distributions of true disk novae, on top
and bottom, show a considerably milder difference.

Fig. 3 suggests a way to estimate the actual M31 bulge--to--disk nova
ratio, just by looking at the bottom--to--top ratio (the BTR) of
apparent bulge novae --- and thus avoiding the issue of the extent to
which the bulge has been searched more thoroughly than the disk.
Because Fig. 2 shows that our model $V$--distribution only fits the
observed $V$--distribution on its bright side, we now confine our
attention to novae having $V < 17$.  The BTR of observed apparent
bulge novae having $V<17$ (see Fig. 1) is $0.83 \pm 0.22$, where the
uncertainty is from $\sqrt(N)$ statistics. The bottom panel of Fig. 4
shows the model distribution of the sky positions of novae having $V <
17$.  As expected, true disk novae show a mild asymmetry with respect
to the major axis, while true bulge novae are strongly concentrated to
the top.  The top panel of Fig. 4 is for an adopted bulge--to--disk
ratio of nine, instead of 1/2, {\sl i.e.}, for the case in which M31
novae are overwhelmingly from the bulge.  Fig. 5 is like Fig. 4, but
showing enlarged views of the bulge.  For the model bulge--to--disk
ratio of 1/2 (bottom panel), the BTR of apparent bulge novae is 0.57.
For the bulge--dominated case (top panel) the BTR ratio of apparent
bulge novae is only 0.25, and the disagreement with the sky positions
of observed novae having $V<17$ (Fig. 1) is obvious.

\section {Discussion}
 
We have found that the assumption that M31 novae come overwhelmingly
from the bulge produces results that are inconsistent with
observation.  Instead, adopting an M31 disk--to--bulge nova ratio that
is like the M31 disk--to--bulge mass ratio produces results that our
acceptable.  This would be consistent with the proposition that the
nova rate per unit K--band luminosity is approximately constant
(Ciardullo et al. 1990; Shafter et al. 1996).

If we take our simple model literally we can go further and derive the
M31 bulge--to--disk nova ratio that actually reproduces the observed
BTR of $0.83 \pm 0.22$.  Fig. 6 shows the dependence of the model BTR
on the percentage of true bulge novae, for three different degrees of
dustiness --- our standard case as described by eqns (1) and (2);
twice as dusty; and half as dusty.  For our standard dust model, even
zero percent bulge novae yields a BTR that is not quite as high as the
observed one; within the statistical uncertainty of the observed BTR
the upper limit on the percentage of bulge novae is 25 percent, {\sl
i.e.,} a bulge--to--disk nova ratio of 1/3.  This would be consistent
with the proposition that young populations are better at producing
novae than old populations (Della Valle et al. 1994; Yungelson et
al. 1997).  However, Fig. 6 shows that should M31 be only half as
dusty as we have assumed (cf. Han 1996), then our upper limit on the
percentage of bulge novae would be 63 percent, {\sl i.e.} a
bulge--to--disk nova ratio of 1.7.  In view of the statistical
uncertainties associated with the observed BTR, and with our simple
model, it probably would be premature to draw any conclusion other
than that the M31 bulge--to--disk nova ratio is at least as low as the
M31 bulge--to--disk mass ratio.

Now, in order to refine our knowledge of the spatial distribution of
novae in M31, what is needed is a carefully controlled search for
novae that includes parts of the disk that are unambiguously outside
the bulge.  It is interesting that of eight M31 novae that were
discovered in a recent search by Sharov \& Alksnis (1996), only three
qualify as apparent bulge novae.  As we completed this study we
learned that another major search for novae in the disk of M31 is
planned (A. Shafter, private communication).

We are grateful to Eddie Baron, Darrin Casebeer, Dean Richardson, and
Lev Yungelson for discussions, and to Allen Shafter for
correspondence.  This work has been supported by NSF grants AST
9417102 and 9417242.

\clearpage

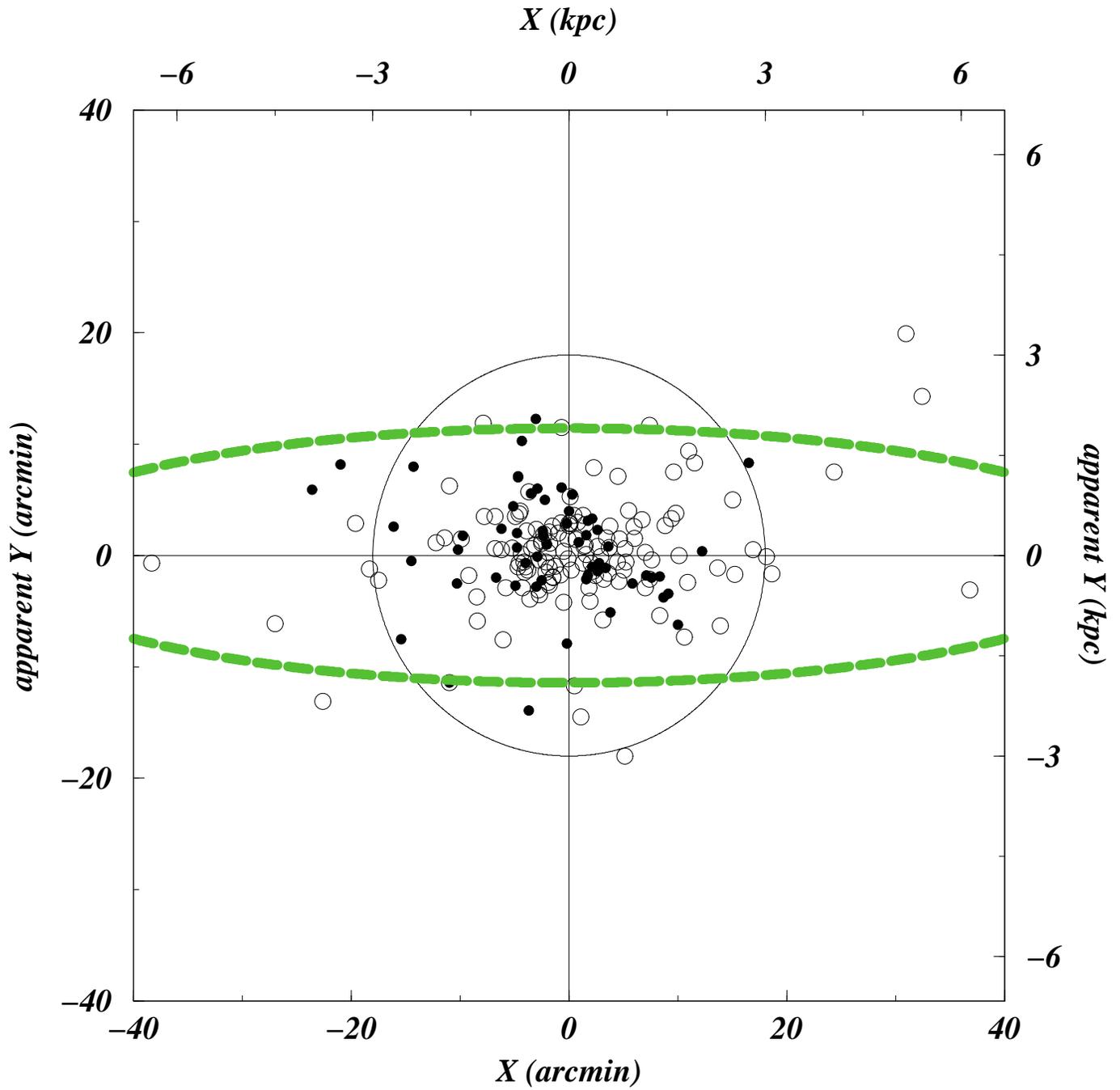
\begin{figure}

\figcaption{Sky positions of 191 novae observed in M31, for
which estimates of peak $V$ are available.  The M31 major and minor
axes are along the X and Y axes, respectively.  Novae within the
18--arcminute ($\simeq 3$ kpc) circle are ``apparent bulge novae".
The ellipse represents a circle in the inclined disk, of radius 8.8
kpc.  Open and filled circles denote novae having $V < 17$ and $V >
17$, respectively.
\label{fig1}}
\end{figure}

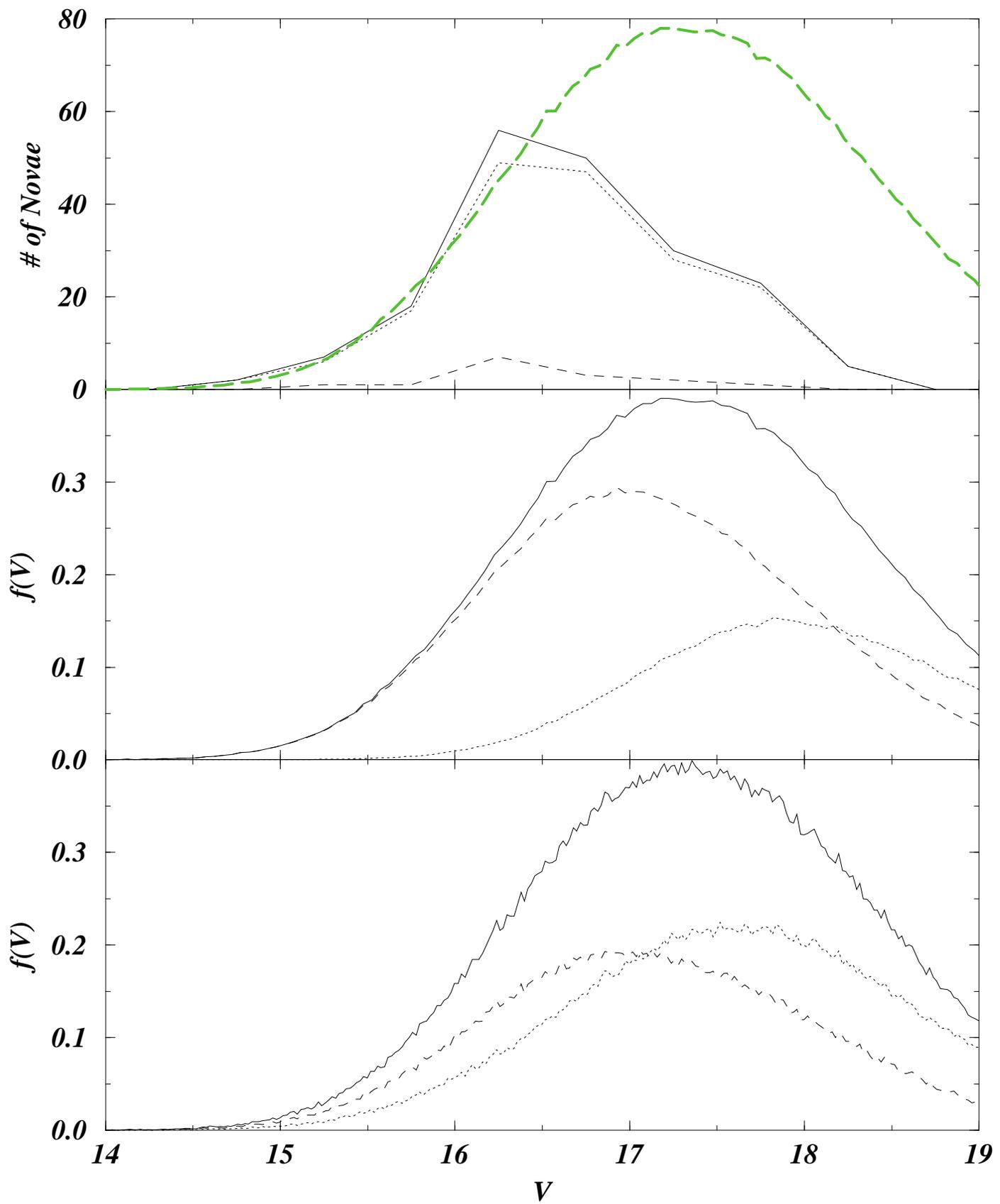
\begin{figure}
\figcaption{({\sl top}): the $V$--distribution of observed
novae in M31; the long--dashed line is for apparent disk novae, the
short--dashed line is for apparent bulge novae, and the solid line is
the sum of the two.  The highest curve is our model
$V$--distribution. ({\sl middle}): The model $V$--distributions for
true disk novae (long--dashed line), true bulge novae (short--dashed
line), and their sum.  ({\sl bottom}): like the middle panel but for
apparent disk and bulge novae.
\label{fig2}}
\end{figure}

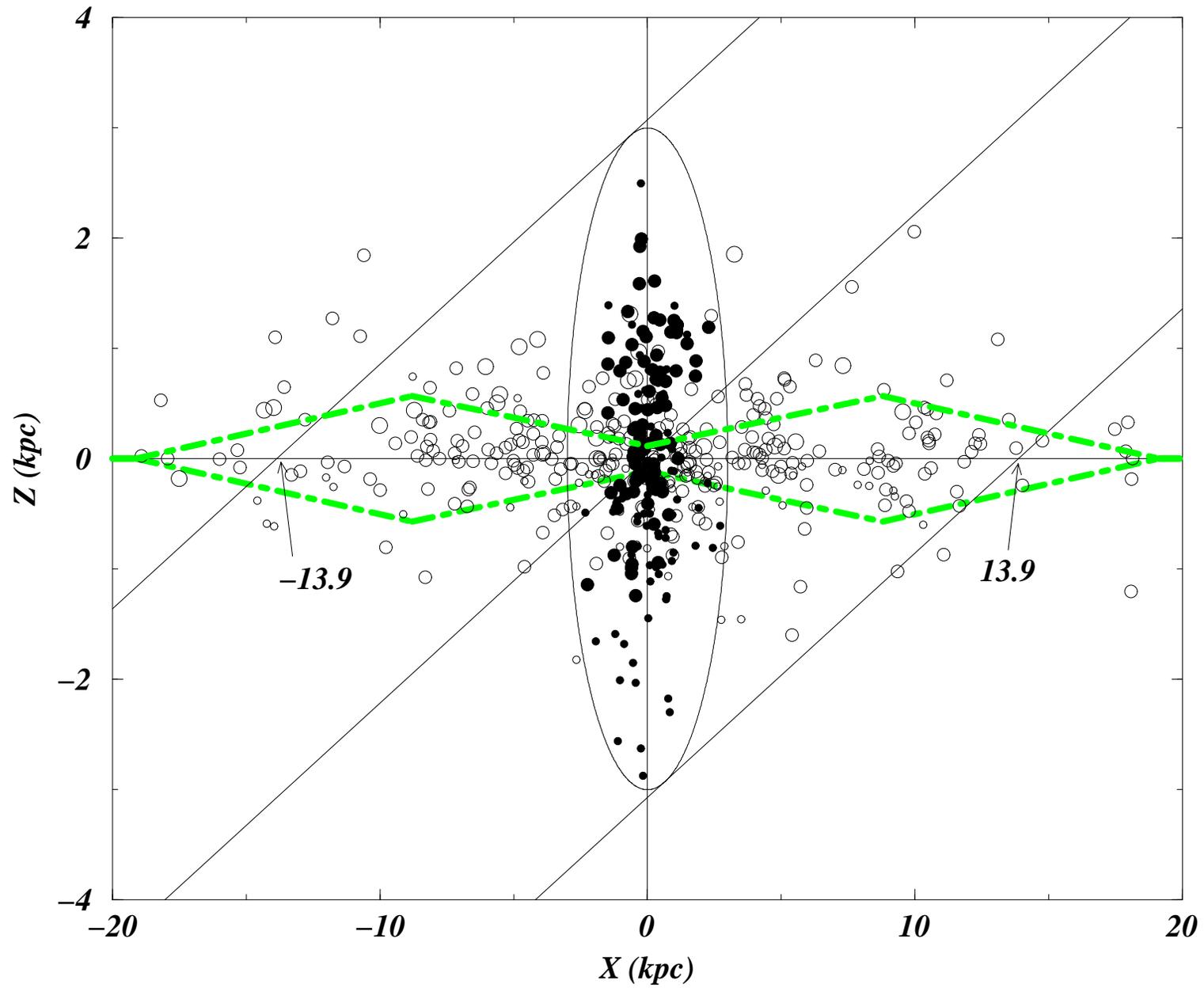
\begin{figure}

\figcaption{A side view of the model spatial distribution of
novae in M31. For clarity, the vertical scale is expanded by a factor
of five.  Our line of sight is from the upper right.  Filled and open
circles denote true bulge and true disk novae, respectively.  Large,
medium, and small symbols denote $V<16, 16 \le V \le 18$, and $V \ge 18$,
respectively.  The widths of the diamond--shaped figures indicate the
adopted radial dependence of the dust density.  From our vantage
point, true bulge novae in the bottom of the bulge tend to be
extinguished by dust that is located well out in the disk.
\label{fig3}}
\end{figure}

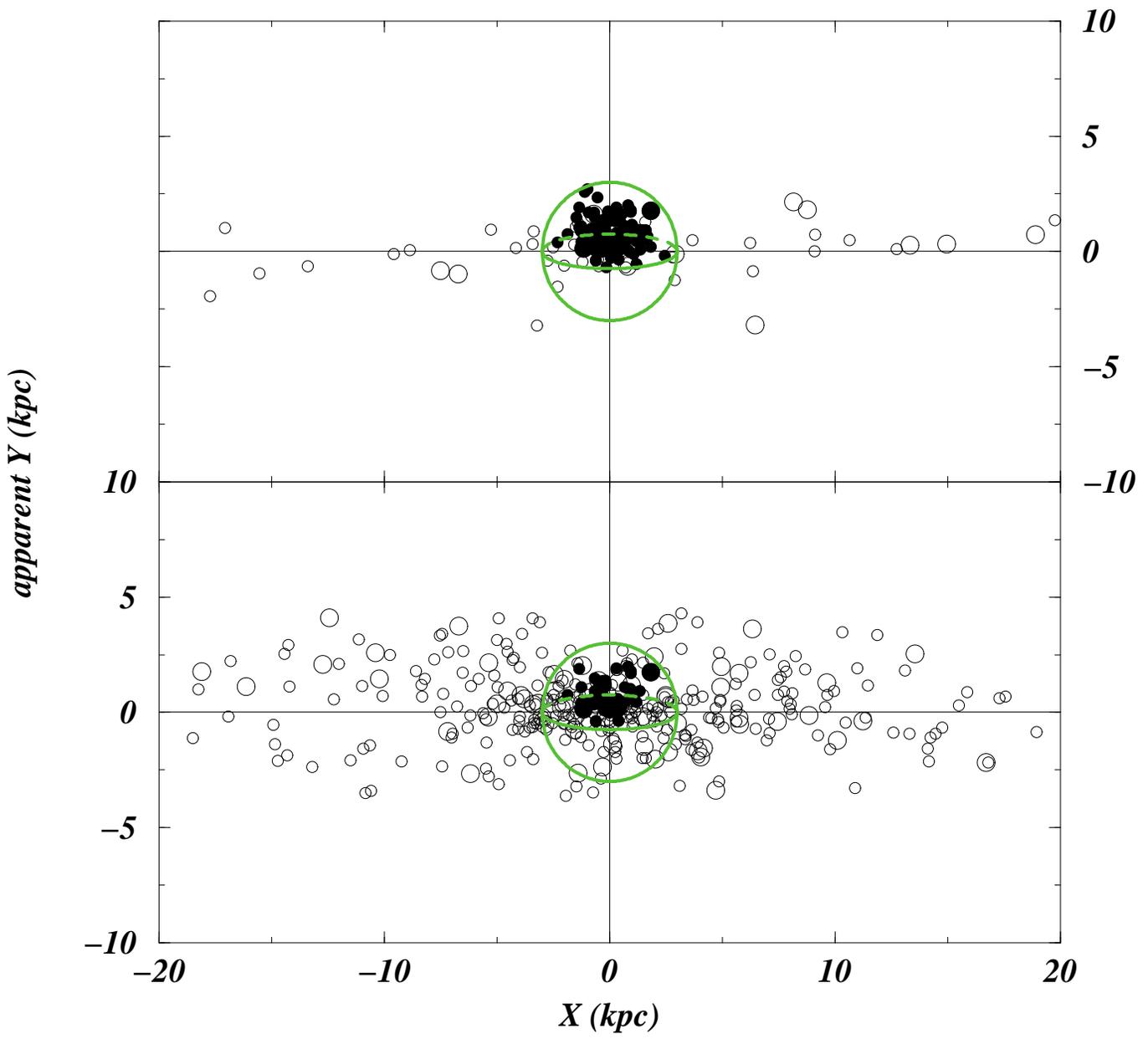
\begin{figure}

\figcaption{{\sl bottom}: the sky positions of model M31 novae
having $V < 17$, for our standard bulge--to--disk nova ratio of 1/2.
Filled and open symbols denote true bulge and true disk novae, and
large and small symbols denote $V<16$ and $V \ge 16$. {\sl top}: like the
bottom panel, but for a bulge--to--disk nova ratio of nine.
\label{fig4}}
\end{figure}

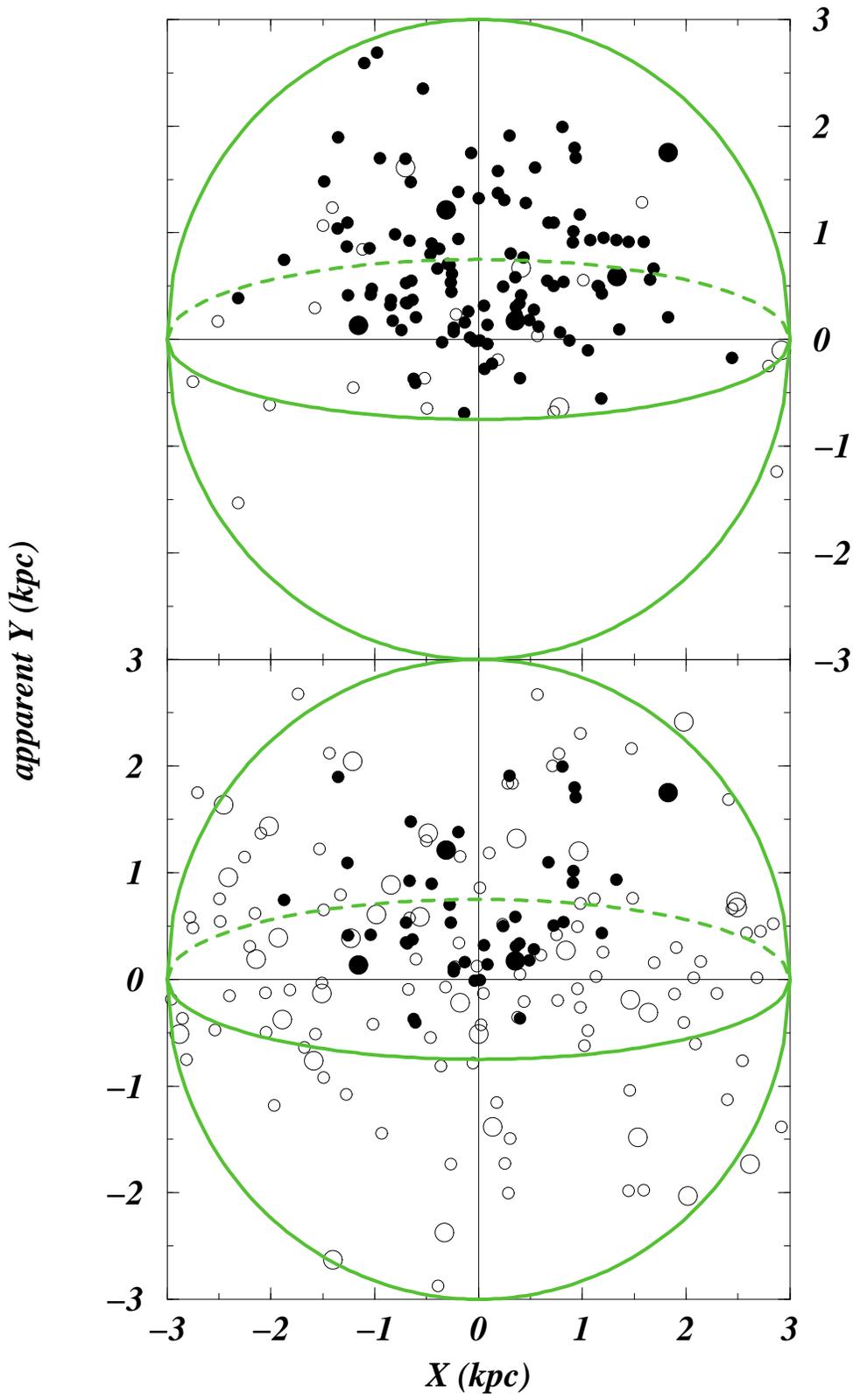
\begin{figure}

\figcaption{Like Fig. 4, but with an expanded view of the bulge.
\label{fig5}}
\end{figure}
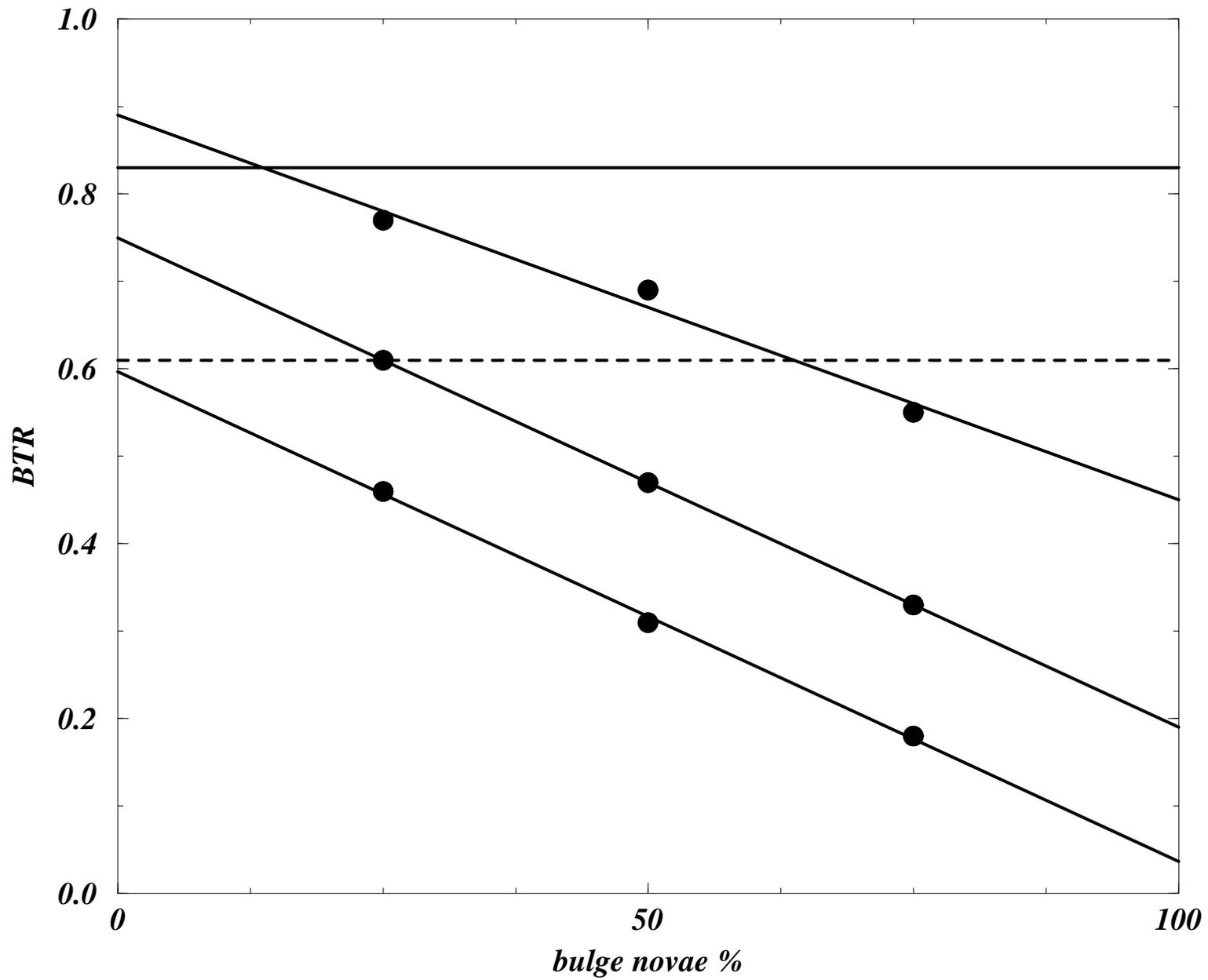
\begin{figure}

\figcaption{The model BTR is plotted against the percentage of
bulge novae, for our standard dust model (central slanted line), for
twice as dusty (lower slanted line), and half as dusty (upper slanted
line).  The solid horizontal line represents the observed BTR of 0.83
and the dashed horizontal line represents the statistical lower limit
of the observed BTR, 0.61.
\label{fig6}}
\end{figure}

\end{document}